\documentclass[10pt,letterpaper]{article}
\usepackage{opex3}
\usepackage{epstopdf}
\usepackage{amssymb}
\usepackage{soul, color}
\usepackage{ulem}
\usepackage{cite}
\newcommand{\ket}[1]{\vert#1\rangle}

\begin{document}

%% NOTE: TITLE PAGE & TOC NOT USED FOR MANUSCRIPT SUBMISSIONS %%
%\title{Modeling a Measurement-Device-Independent Quantum Key Distribution System}

%\vskip4pc

%\tableofcontents
%\clearpage
%% NO TITLE PAGE FOR OPEX SUBMISSIONS %%

%% START HERE
%%%%%%%%%%%%%%%%%% title page information %%%%%%%%%%%%%%%%%%
\title{Modeling a measurement-device-independent quantum key distribution system}

\author{P. Chan,$^{1,2}$ J. A. Slater,$^{1,3}$ I. Lucio-Martinez,$^{1,3}$ A. Rubenok,$^{1,3}$ W.~Tittel$^{1,3}$}

\address{$^1$Institute for Quantum Science \& Technology, University of Calgary, 2500 University Drive NW, Calgary, Alberta T2N 1N4, Canada}
\address{$^2$Department of Electrical \& Computer Engineering, University of Calgary, 2500 University Drive NW, Calgary, Alberta T2N 1N4, Canada}
\address{$^3$Department of Physics \& Astronomy, University of Calgary, 2500 University Drive NW, Calgary, Alberta T2N 1N4, Canada}

\email{wtittel@ucalgary.ca} %% email address is required



\begin{abstract} 
We present a detailed description of a widely applicable mathematical model for quantum key distribution (QKD) systems implementing the measurement-device-independent (MDI) protocol. The model is tested by comparing its predictions with  data taken using a proof-of-principle, time-bin qubit-based QKD system in a secure laboratory environment (i.e. in a setting in which eavesdropping can be excluded). The good agreement between the predictions and the experimental data allows the model to be used to optimize mean photon numbers per attenuated laser pulse, which are used to encode quantum bits. This in turn allows optimization of secret key rates of existing MDI-QKD systems, identification of rate-limiting components, and  projection of future performance. In addition, we also performed measurements over deployed fiber, showing that our system's performance is not affected by environment-induced perturbations.
\end{abstract}

 % REPLACE WITH CORRECT OCIS CODES FOR YOUR ARTICLE

%%%%%%%%%%%%%%%%%%%%%%% References %%%%%%%%%%%%%%%%%%%%%%%%%

%%%%%%%%%%%%%%%%%%%%%%%%%%  body  %%%%%%%%%%%%%%%%%%%%%%%%%%
\section{Introduction \label{sec:introduction}}

From the first proposal in 1984 to now, the field of quantum key distribution (QKD) has evolved significantly~\cite{Gisin2002, Scarani2009}. For instance, experimentally, systems delivering key at Mbps rates~\cite{Dixon2010} as well as key distribution over more than 100 km~\cite{Stucki2009, Schmitt2007} have been reported. From a theoretical perspective, efforts aim at developing QKD protocols and security proofs with minimal assumptions about the devices used~\cite{Masanes2011}. Of particular practical importance are two recently developed protocols that do not require trusted single photon detectors (SPDs)~\cite{Lo2012, Braunstein2012}. 
 One of these, the so-called measurement-device-independent QKD (MDI-QKD) protocol, has already been  implemented experimentally~\cite{Rubenok2013, Liu2012, daSilva2012,Tang2013}. 
Hence, it is foreseeable that it will play an important role in the future of QKD, and it is thus important to understand the interplay between experimental imperfections (which will always remain in real systems) and system performance to maximize the latter.

In this work, we derive a widely applicable mathematical model describing systems that implement the MDI-QKD protocol. The model is based on facts about our \cite{Rubenok2013}, and other existing experimental setups \cite{Liu2012, daSilva2012,Tang2013}, and takes into account carefully characterized imperfect state preparation, loss in the quantum channel, as well as limited detector efficiency and noise. It is tested by comparing its predictions with data taken with a proof-of-principle QKD system~\cite{Rubenok2013} employing time-bin qubits and implemented in a laboratory environment. Our model, which contains no free parameter, reproduces the experimental data within statistical uncertainties over three orders of magnitude of a relevant parameter. The excellent agreement allows optimizing central parameters that determine secret key rates, such as mean photon numbers used to encode qubits, and to identify rate-limiting components for future system improvement. In addition, we also find that the model accurately reproduces experimental data obtained over deployed fibers, showing that our system minimizes environment-induced perturbation to quantum key distribution in real-world settings.

This paper is organized in the following way: In section~\ref{sec:sidechannels} we detail some of the side-channel attacks (i.e. attacks exploiting incorrect assumptions about the working of QKD devices) proposed so far and review technological  countermeasures. In section~\ref{sec:MDI} we briefly describe the MDI-QKD protocol, which instead exploits fundamental quantum physical laws to render the most important of these attacks useless. Our model of MDI-QKD systems is presented in section~\ref{sec:model}. This section is followed by an in-depth account of experimental imperfections that affect MDI-QKD performance and a description of how we characterized them in our system (section~\ref{sec:characterization}). Section~\ref{sec:verification} shows the results of the comparison between modelled and measured quantities, and section~\ref{sec:optimization} details how to optimize the performance of our MDI-QKD system using the model. Finally, we conclude the article in section~\ref{sec:conclusion}.

\begin{figure}[h!]
\centering \includegraphics[width=13cm]{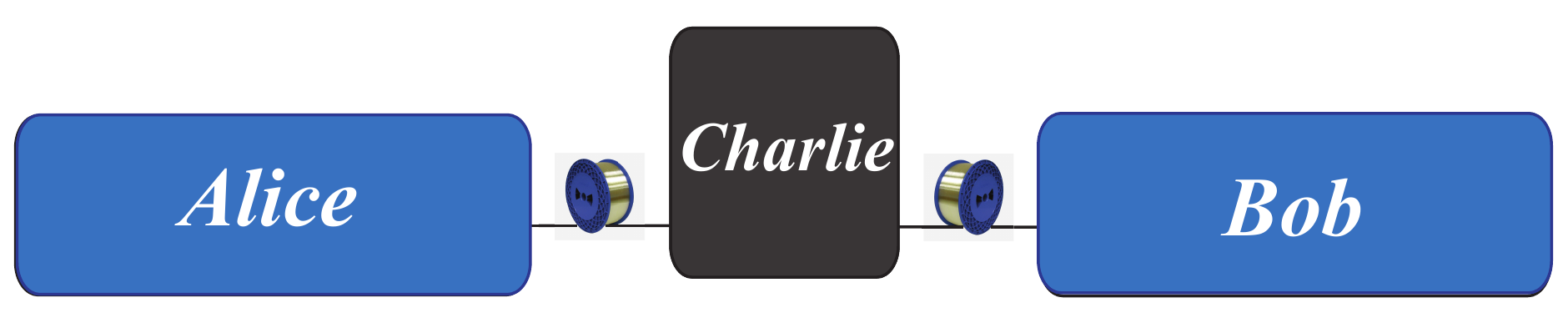}
	\begin{center}
		\caption{\label{fig:schematic} Schematics for MDI-QKD. Charlie facilitates the key distribution between Alice and Bob without being able to learn the secret key. }
	\end{center}
\end{figure}

\section{Side-channel attacks \label{sec:sidechannels}}
A healthy development of QKD requires investigating the vulnerabilities of QKD implementations in terms of potential side-channel attacks. Side-channels in QKD are channels over which information about the key may leak out unintentionally. One of the first QKD side-channel attacks proposed was the photon number splitting (PNS) attack~\cite{Brassard2000} in which the eavesdropper, Eve, exploits the fact that attenuated laser pulses sometimes include more than one photon to obtain information about the key. This attack can be detected if the decoy state protocol~\cite{Hwang2003, Lo2005, Wang2005} is implemented. In the decoy state protocol, Alice varies the mean photon number per pulse in order to allow her and Bob to distill the secret key only from information stemming from single photon emissions.  More proposals of side-channel attacks followed, including the Trojan-horse attack~\cite{Gisin2006}, for which the countermeasure is an optical isolator~\cite{Gisin2006}, and the phase remapping attack~\cite{Fung2007}, for which the countermeasure is phase randomization~\cite{Fung2007}.
Later on, attacks that took advantage of SPD vulnerabilities were also proposed and demonstrated~\cite{Lamas2007, Zhao2008, Wiechers2010, Lydersen2010}. For example, the time-shift attack~\cite{Zhao2008} exploits a difference in the quantum efficiencies of the SPDs used in a QKD system. This attack can be prevented by actively selecting one of the two bases for the projection measurement, as well as by monitoring the temporal distribution of  photon detections~\cite{Zhao2008}. Another example is the detector blinding attack~\cite{Lydersen2010} in which the eavesdropper uses high intensity pulses to modify the performance (i.e. blind) the SPDs. It can be detected by monitoring the intensity of light at the entrance of Bob's devices with a photodiode~\cite{Lydersen2010, Avoiding1, Avoiding2}.  Nevertheless, due to its power, the blinding attack is currently of particular concern.

It is important to mention that open side-channels do not necessarily compromise the security of the final key if the information that Eve may have obtained through an attack is properly removed during privacy amplification. However, as technological fixes (as discussed above) or additional privacy amplification can only thwart known attacks, it is important to develop and implement protocols that use a minimum number of assumptions about the devices used to implement the protocol. An important example is the measurement-device-independent QKD protocol, which we will introduce in the next section.

\section{The measurement-device-independent quantum key distribution protocol\label{sec:MDI}}
The MDI-QKD protocol is a time-reversed version of entanglement-based QKD. In this protocol, the users, Alice and Bob, are each connected to Charlie, a third party, through a quantum channel, e.g. optical fiber (see Fig.~\ref{fig:schematic}). In the ideal version, the users have a source of single photons that they prepare randomly in one of the BB84 qubit states \cite{BB84} $\ket{0}$, $\ket{1}$, $\ket{+}$ and $\ket{-}$, where $\ket{\pm} = 2^{-1/2}(\ket{0} \pm \ket{1})$. The qubits are sent to Charlie where the SPDs are located. Charlie performs a partial Bell state measurement (BSM) through a 50/50 beam splitter and then announces the events for which the measurement resulted in a projection onto the $\ket{\psi^-}= 2^{-1/2} (\ket{0}_A\ket{1}_B - \ket{1}_A\ket{0}_B)$ state. Alice and Bob then publicly exchange information about the used bases (z, spanned by $\ket{0}$ and $\ket{1}$, or x, spanned by $\ket{+}$ and $\ket{-}$). Associating quantum states with classical bits (e.g. $\ket{0},\ket{-}\equiv~$0, and $\ket{1},\ket{+}\equiv$~1) and keeping only events in which Charlie found $\ket{\psi^-}$ and they picked the same basis, Alice and Bob now establish anti-correlated key strings. (Note that a projection of two photons onto $\ket{\psi^-}$ indicates that the two photons, if prepared in the same basis, must have been in orthogonal states.) Bob then flips all his bits, thereby converting the anti-correlated strings into correlated ones. 
Next, the so-called \textit{x-key} is formed out of all key bits for which Alice and Bob prepared their photons in the x-basis; its error rate is used to bound the information an eavesdropper may have acquired during photon transmission. Furthermore, Alice and Bob form the \textit{z-key} out of those bits for which both picked the z-basis.  Finally, they perform error correction and privacy amplification\cite{Gisin2002,Scarani2009} to the \textit{z-key},  which results in the secret key.

The advantage of the MDI-QKD protocol over conventional prepare-and-measure or entangled photon-based QKD protocols is that, in the case of Charlie performing an ideal (partial) BSM as described above, detection events are uncorrelated with the final secret key bits. This is because a projection onto $\ket{\psi^-}$ only indicates that Alice and Bob sent orthogonal states, but does not reveal who sent which state. As a result, Charlie (or Eve) is unable to gain any information about the key from passively monitoring the detectors. Furthermore, a measurement that is different from the ideal BSM leads to an increased error rate and thus to a smaller, but still secret, key once privacy amplification has been applied. Notably, it does not matter wether the difference is due to experimental imperfections or to an eavesdropper (possibly Charlie himself) trying to gather information about the states that Alice and Bob sent by replacing or modifying the measurement apparatus. Hence, all detector side channels are closed in MDI-QKD.

In the ideal scenario introduced above, Alice and Bob use single photon sources to generate qubits. However, it is possible to implement the protocol using light pulses attenuated to the single photon level. Indeed, as in prepare-and-measure QKD, randomly varying the mean photon number of photons per attenuated light pulse between a few different values (so-called decoy and signal states) allows making the protocol practical while protecting against a possible PNS attack~\cite{Lo2012,Wang2013}.
The secret key rate is then given by~\cite{Lo2012}:

\begin{equation}
S=Q_{11}^z\big (1-h_2(e_{11}^x)\big ) - Q_{\mu\sigma}^z f h_2(e_{\mu\sigma}^z ),
\label{eq:secret_key_rate}
\end{equation}

\noindent where $h_2$ is the binary entropy function, $f$ indicates the error correction efficiency, $Q$ indicates the gain (the probability of a projection onto $\ket{\psi^-}$ per emitted pair of pulses~\cite{note}) and $e$ indicates error rates (the ratio of erroneous to total projections onto $\ket{\psi^-}$). Furthermore, the superscripts, $x$ or $z$, denote if gains or error rates are calculated for qubits prepared in the x- or the z-basis, respectively. Similarly, the subscripts, $\mu$ and $\sigma$, show that the quantity under concern is calculated or measured for pulses with mean photon number $\mu$ (sent by Alice) and $\sigma$ (sent by Bob), respectively. Finally, the subscript $11$ indicates quantities stemming from detection events for which the pulses emitted by Alice and Bob contain only one photon each. Note that $Q_{11}$ and $e_{11}$ cannot be measured; their values must be bounded using either a decoy state method, or employing qubit tagging \cite{Brassard2000}. However, the latter yields smaller key rates and distances than the former.

Shortly after the original proposal \cite{Lo2012}, a practical decoy state protocol for MDI-QKD was proposed~\cite{Wang2013}. It requires Alice and Bob to randomly pick mean photon numbers between two decoy states and a signal state. One of the decoy states must have a mean photon number lower than the signal state, while the other one must be vacuum. A finite number of decoy states results in a lower bound for $Q_{11}^{x,z}$ and an upper bound for $e_{11}^x$, which in turn gives a lower bound for the secret key rate in Eq.~(\ref{eq:secret_key_rate}). We will elaborate more on decoy states in section~\ref{sec:decoy}.

\section{The model\label{sec:model}} 
Our model takes into account imperfections present in a typical QKD system. Regarding the sources, located at Alice and Bob, we take into account imperfect preparation of the quantum  state of each photon. Furthermore, we consider transmission loss of the links between Alice and Charlie, and Bob and Charlie. And finally, concerning the measurement apparatus at Charlie's, we consider imperfect projection measurement stemming from non-maximum quantum interference on Charlie's beam splitter, detector noise such as dark counts and afterpulsing, and limited detector efficiency. See also~\cite{Xu2013} for another model describing MDI-QKD performance, but with a more restrictive set of imperfections and not yet tested against actual experimental data.

In the following paragraphs we present a detailed description of our model. It relies on the assumption of phase randomized laser pulses at Charlie's. While Alice and Bob generate coherent states in our proof-of-principle setup, this assumption is correct as the long fibres used to connect Alice and Bob with Charlie introduce random global phase variations~(we will discuss the impact of the lack of phase randomization at Alice's and Bob's on the security of distributed keys in section \ref{sec:conclusion}). We note that, in order to facilitate explanations, we have adopted the terminology of time-bin encoding. However, our model is general and can also be applied to MDI-QKD systems implementing other types of encoding \cite{daSilva2012}.

\subsection{State preparation \label{sec:state}}
In the MDI-QKD protocol, Alice and Bob derive key bits whenever Charlie announces a projection onto the $\ket{\psi^-}$ Bell state. We model the probability of a $\ket{\psi^-}$ projection for various quantum states of photons emitted by Alice and Bob as a function of the mean photon number per pulse ($\mu$ and $\sigma$, respectively) and transmission coefficients of the fiber links ($t_A$ and $t_B$, respectively). We consider photons in qubit states described by:

\begin{equation}
\ket{\psi}=\frac{1}{\sqrt{1+2b^{x,z}}}\left(\sqrt{m^{x,z}+b^{x,z}}\ket{0} + e^{i\phi^{x,z}}\sqrt{1-m^{x,z}+b^{x,z}}\ket{1}\right)
\label{eqn:photon_state}
\end{equation}

\noindent where $\ket{0}$ and $\ket{1}$ denote orthogonal modes (i.e. early and late temporal modes assuming time-bin qubits), respectively.  Note that $\ket{\psi}$ describes any pure state~\cite{note2} and the presence of the  $m^{x,z}$ and $b^{x,z}$ terms in Eq.~(\ref{eqn:photon_state}), as opposed to using only one parameter, is motivated by the fact that they model different experimentally characterizable imperfections. In the ideal case,  $m^{z}$ $\in [0,1]$  for photon preparation in the z-basis (in this case, the value of $\phi^z$ is irrelevant), $m^{x}=\frac{1}{2}$ and $\phi^{x}\in[0,\pi]$ for the x-basis, and $b^{x,z}=0$ for both bases. Imperfect preparation of photon states is modelled by using non-ideal $m^{x,z}$, $\phi^{x,z}$ and $b^{x,z}$ for Alice and Bob. The parameter $b^{x,z}$ is included to represent the background light emitted and modulated by an imperfect source.  Furthermore, in principle, the various states generated by Alice and Bob could have differences in other degrees of freedom (i.e. polarization, spectral, spatial, temporal modes). This is not included in Eq.~(\ref{eqn:photon_state}), but would be reflected in a reduced quality of the BSM, which will be discussed below. 

\subsection{Conditional probability for projections onto $\ket{\psi^-}$ \label{sec:interference}}
A projection onto $\ket{\psi^-}$ occurs if one of the SPDs after Charlie's 50/50 beam splitter signals a detection in an early time-bin (a narrow time interval centered on the arrival time of photons occupying an early temporal mode) and the other detector signals a detection in a late time-bin (a narrow time-interval centered on the arrival time of photons occupying a late temporal mode). Note that, in the following paragraphs, this is the desired detection pattern we search for when modeling possible interference cases or noise effects. Also, note that we assume that Charlie's two single-photon detectors have identical properties. A deviation from this approximation does not open a potential security loophole (in contrast to prepare-and-measure and entangled photon based QKD), as all detector side-channel attacks are removed in MDI-QKD.

We build up the model by first considering the probabilities that particular outputs from the beam splitter (at Charlie's) will generate the detection pattern associated with a projection onto $\ket{\psi^-}$.  The outputs are characterized by the number of photons per output port as well as their joint quantum state. The probabilities for each of the possible outputs to occur can then be calculated based on the inputs to the beam splitter (characterized by the number of photons per input port and their quantum states, as defined in Eq.~(\ref{eqn:photon_state})). Note that for the simple cases of inputs containing zero or one photon (summed over both input modes), we calculate the probabilities leading to the desired detection pattern directly, i.e. without going through the intermediate step of calculating outputs from the beam splitter. Finally, the probability for each input to occur is calculated based on the probability for Alice and Bob to send attenuated light pulses containing exactly $i$ photons, all in a state given by Eq.~(\ref{eqn:photon_state}). The probability for a particular input to occur also depends on the transmissions of the quantum channels, $t_A$ and $t_B$. We note that this model considers up to three photons incident on the beam splitter. This is sufficient as, in the case of heavily attenuated light pulses and lossy transmission, higher order terms do not  contribute significantly to projections onto $\ket{\psi^-}$. However, we limit the following description to two photons at most: the extension to three is lengthy but straightforward and follows the methodology presented for two photons.

\paragraph{Detector noise  \label{sec:detector}}
Let us begin by considering the simplest case in which no photons are input into the beam splitter.  In this case, detection events can only  be caused by detector noise. We denote the probability that a detector indicates a spurious detection as $P_{n}$.  Detector noise stems from two effects: dark counts and afterpulsing~\cite{Stucki2001}.  Dark counts represent the base level of noise in the absence of any light, and we denote the probability that a detector generates a  dark count per time-bin as $P_{d}$.  Afterpulsing is an additional noise source produced by the detector as a result of prior detection events.  The probability of afterpulsing depends on the total count rate, hence we denote the afterpulsing probability per time-bin as $P_a$, which is a function of the mean photon number per pulse from Alice and Bob ($\mu$ and $\sigma$), the transmission of the channels ($t_A$ and $t_B$) and the efficiency of the detectors ($\eta$) located at Charlie (see below for afterpulse characterization). The total probability of a noise count in a particular time-bin is thus $P_{n} = P_{d} + P_{a}$. All together, we find the probability for generating the detection pattern associated with a projection onto the $\ket{\psi^-}$-state, conditioned on having no photons at the input, specified by ``in", of the beam splitter, to be :

\begin{equation}
P(\ket{\psi^-}| \mbox{0 photons, in}) =  P(\ket{\psi^-}| \mbox{0 photons, out})= 2P_{n}^2,\\
\label{eqn:0_photons}
\end{equation}
\noindent
Here and henceforward, we have ignored the multiplication factor (1-$P_{n}) \sim 1$~\cite{note3}, which indicates the probability that a noise event did not occur in the early time-bin (this is required in order to see a detection during the late time-bin assuming detectors with recovery time larger than the separation between the $\ket{0}$ and $\ket{1}$ temporal modes). Note that the probability conditioned on having no photons at the inputs of the beam splitter equals the one conditioned on having no photons at the outputs (specified in Eq.~(\ref{eqn:0_photons}) by the conditional ``out").

\paragraph{One-photon case  \label{sec:one}}
Next, we consider the case in which a single photon arrives at the beam splitter. To generate the detection pattern associated with $\ket{\psi^-}$, either the photon must  be detected and a noise event must occur in the other detector in the opposite time-bin, or, if the  photon is not detected, two noise counts must occur as in Eq.~(\ref{eqn:0_photons}).  We find

\begin{equation}
P(\ket{\psi^-} | \mbox{1 photon, in}) = \eta P_{n} + (1-\eta)P(\ket{\psi^-} | \mbox{0 photons, out}),
\label{eqn:1_photon}
\end{equation}
\noindent
where $\eta$ denotes the probability to detect a photon that occupies an early (late) temporal mode during an early (late) time-bin (we assume $\eta$ to be the same for both detectors). 

\paragraph{Two-photon case  \label{sec:two}}
We now consider detection events stemming from two photons entering the beam splitter. The possible outputs can be broken down into three cases.  In the first case, both photons exit the beam splitter in the same output port and are directed to the same detector. This yields only a single detection event, even if the photons are in different temporal modes (the latter is due to detector dead time. Note that as our model calculates detections in units of bits per gate, modeling a dead-time free detector is straightforward.). The probability for Charlie to declare a projection onto $\ket{\psi^-}$ is then

\begin{eqnarray}
P(\ket{\psi^-} | \mbox{2 photons, 1 spatial mode, out}) =\nonumber \\
\hspace{3mm} (1-(1-\eta)^2)P_{n} + (1-\eta)^2P(\ket{\psi^-} | \mbox{0 photons, out}).
\label{eqn:2_photon_1_spatial}
\end{eqnarray}

In the second case, the photons are directed towards different detectors and occupy the same temporal mode. Hence, to find detections in opposite time-bins in the two detectors, at least one photon must not be detected. This leads to

\begin{eqnarray}
P(\ket{\psi^-} | \mbox{2 photons, 2 spatial modes, 1 temporal mode, out}) = \nonumber \\
\hspace{3mm} 2\eta(1-\eta)P_{n} + (1-\eta)^2P(\ket{\psi^-} | \mbox{0 photons, out}).
\label{eqn:2_photon_1_temporal}
\end{eqnarray}

In the final case, both photons occupy different spatial  as well as temporal modes.  In contrast to the previous case, a projection onto $\ket{\psi^-}$ can now also originate from the detection of both photons. This leads to

\begin{eqnarray}
P(\ket{\psi^-} | \mbox{2 photons, 2 spatial modes, 2 temporal modes, out}) =  \nonumber \\
\hspace{3mm} \eta^2 + 2\eta(1-\eta)P_{n} + (1-\eta)^2P(\ket{\psi^-} | \mbox{0 photons, out}).
\label{eqn:2_photon_2_temporal}
\end{eqnarray}

In order to find the probability for each of these three two-photon outputs to occur, we must examine two-photon inputs to the beam splitter. We note that it is possible for the two photons to be subject to a two-photon interference effect (known as photon bunching) when impinging on the beam splitter. As this quantum interference can lead to an entangled state between the output modes, the calculation must proceed with quantum mechanical operators. We consider three cases:  two photons arrive at the same input of the beam splitter, one photon arrives at each input of the beam splitter and the two photons 
are distinguishable, and one photon arrives at each input of the beam splitter and the two photons are indistinguishable. For ease of analysis, we first introduce some notation:

\begin{eqnarray}
p^{x,z}(0,0) & \equiv & (m^{x,z}_1+b^{x,z}_1)(m^{x,z}_2+b^{x,z}_2) \nonumber\\
p^{x,z}(0,1) & \equiv & (m^{x,z}_1+b^{x,z}_1)(1 - m^{x,z}_2+b^{x,z}_2)\nonumber\\
p^{x,z}(1,0) & \equiv & (1-m^{x,z}_1+b^{x,z}_1)(m^{x,z}_2+b^{x,z}_2)\nonumber\\
p^{x,z}(1,1) & \equiv & (1-m^{x,z}_1+b^{x,z}_1)(1-m^{x,z}_2+b^{x,z}_2)\nonumber\\
b^{x,z}_{norm} & \equiv & 1+ 2b^{x,z}_1 + 2b^{x,z}_2 +4b^{x,z}_1b^{x,z}_2
\label{eqn:p_definitions}
\end{eqnarray}

\noindent
where $b^{x,z}_{1,2}$ and $m^{x,z}_{1,2}$ are the parameters introduced in Eq.~(\ref{eqn:photon_state}); the subscripts label the photon (one or two) whose state is specified by the parameters.  Furthermore, $p^{x,z}(i,j)$ is proportional to finding photon one before the beam-splitter in temporal mode $i$ and photon two in  temporal mode $j$, where $i,j \in[0,1]$. Finally, $b^{x,z}_{norm}$ is a normalization factor.

First, considering the situation in which the two photons impinge from the same input on the beam splitter, one has the state

\begin{equation}
\ket{\psi_{input}} = \left( \frac{1}{\sqrt{1+2b^{x,z}}} \left(\sqrt{m^{x,z}+b^{x,z}} \ \hat{a}^\dagger (0)+ e^{i\phi^{x,z}}\sqrt{1-m^{x,z}+b^{x,z}} \ \hat{a}^\dagger (1)\right) \right)^{\otimes 2} \ket{vac}, 
\end{equation}
\noindent
where $\hat{a}^\dagger (0)$ and $\hat{a}^\dagger(1)$ are the creation operators for a photon in the $\ket{0}$ or $\ket{1}$ state, respectively. Evolving this state through the standard unitary transformation for a lossless, 50/50 beam splitter, described by $\hat{a}^\dagger \rightarrow (\hat{c}^\dagger + \hat{d}^\dagger)/\sqrt{2}$ (where $\hat{c}^\dagger$ and $\hat{d}^\dagger$ are the two output modes of the beam splitter),  one finds that with probability $1/2$ the two photons exit the beam splitter in the same output port (or spatial mode) and with probability $1/2$ in different ports. Furthermore, with probability $A = [p^{x,z}(0,0) + p^{x,z}(1,1)]/2b^{x,z}_{\mbox{norm}}$ we find the photons in different spatial modes and in the same temporal mode, and with probability $B = [p^{x,z}(0,1) + p^{x,z}(1,0)]/2b^{x,z}_{\mbox{norm}}$ we find the photons in different spatial and temporal modes. By symmetry, we find the same result if the two photons arrive from the other input mode of the beam splitter.

Thus the probability that Charlie finds the desired detection pattern is:
\begin{eqnarray}
P(\ket{\psi^-} | \mbox{2  photons, 1 spatial mode, in} ) = \nonumber \\
\hspace{3mm} \frac{1}{2}P(\ket{\psi^-} | \mbox{2 photons, 1 spatial mode, out}) \nonumber \\
\hspace{3mm} + A \times  P(\ket{\psi^-} | \mbox{2 photons, 2 spatial modes, 1 temporal mode, out}) \nonumber \\
\hspace{3mm} + B \times P(\ket{\psi^-} | \mbox{2 photons, 2 spatial modes, 2 temporal modes, out}).\nonumber\\
\label{eqn:2_photons_1_spatial_modes_non_interfering}
\end{eqnarray}

Second, consider the situation in which the two photons come from different inputs, and are completely distinguishable in some degree of freedom. This can be modelled by starting with the input state

\begin{eqnarray}
	\ket{\psi_{input}} &= \frac{1}{\sqrt{1+2b_1^{x,z}}} \left(\sqrt{m_1^{x,z}+b_1^{x,z}} \ \hat{a}^\dagger (0) + e^{i\phi_1^{x,z}}\sqrt{1-m_1^{x,z}+b_1^{x,z}} \ \hat{a}^\dagger (1)\right) \nonumber \\
			&\otimes \frac{1}{\sqrt{1+2b_2^{x,z}}} \left(\sqrt{m_2^{x,z}+b_2^{x,z}} \ \hat{b}^\dagger (0) + e^{i\phi_2^{x,z}}\sqrt{1-m_2^{x,z}+b_2^{x,z}} \ \hat{b}^\dagger (1)\right)    \ket{vac}, \label{eqn:two_photon_input_state}
\end{eqnarray}
where $\hat{b}^\dagger$ is the creation operator for a photon in the second input mode of the beam splitter. One can then evolve the state with the beam splitter unitary described by $\hat{a}^\dagger \rightarrow (\hat{c}^\dagger + \hat{d}^\dagger)/\sqrt{2}$ (as before) and $\hat{b}^\dagger \rightarrow (-\hat{e}^\dagger + \hat{f}^\dagger)\sqrt{2}$, where $\hat{c}^\dagger$ and $\hat{e}^\dagger$ correspond to the same spatial output  mode but with distinguishability in another degree of freedom, and similarly for the other spatial output mode described by $\hat{d}^\dagger$ and $\hat{f}^\dagger$.  One finds the same result as for the previous case, described by Eq.~(\ref{eqn:2_photons_1_spatial_modes_non_interfering}):

\begin{eqnarray}
P&(\ket{\psi^-} | \mbox{2  photons, 2 spatial modes, non-interfering, in} )\nonumber \\ &= P(\ket{\psi^-} | \mbox{2  photons, 1 spatial mode, in} ) \nonumber \\ 
&\equiv P(\ket{\psi^-} | \mbox{2  photons, non-interfering, in} ).
\label{eqn:2_photons_non_interfering}
\end{eqnarray}
\noindent
The definition reflects that there is no two-photon interference in both cases.

Finally, consider the case in which the two photons impinge from different inputs are indistinguishable, and interfere on the beam splitter. This can be modelled by considering the same input state as in Eq.~(\ref{eqn:two_photon_input_state}), but using a beam splitter unitary described by $\hat{a}^\dagger \rightarrow (\hat{c}^\dagger + \hat{d}^\dagger)/\sqrt{2}$ (as before) and $\hat{b}^\dagger \rightarrow (-\hat{c}^\dagger + \hat{d}^\dagger)/\sqrt{2}$. In this case, the probabilities of finding the outputs  from the beam splitter discussed in Eqs.~(\ref{eqn:2_photon_1_spatial}-\ref{eqn:2_photon_2_temporal})  depend on the difference between the phases $\phi^{x,z}_1$ and $\phi^{x,z}_2$  that specify the states of photons one and two, $\Delta \phi^{x,z} \equiv \phi^{x,z}_1 - \phi^{x,z}_2$. Note that, due to the two-photon interference effect, finding the two photons in different spatial modes and the same temporal mode is impossible. We are thus left with the case of having two photons in the same output port (the same spatial mode), which occurs with probability $C = [p^{x,z}(0,0) + p^{x,z}(1,1) + 0.5(p^{x,z}(0,1) + p^{x,z}(1,0)) + \sqrt{p^{x,z}(0,1)p^{x,z}(1,0)}\cos(\Delta\phi^{x,z})]/b^{x,z}_{norm}$, and the case of having the photons in different temporal and spatial modes, which occurs with probability $D = [0.5(p^{x,z}(0,1) + p^{x,z}(1,0)) -\sqrt{p^{x,z}(0,1)p^{x,z}(1,0)}\cos(\Delta\phi^{x,z})]/b^{x,z}_{norm} $.  This leads to

\begin{eqnarray}
P(\ket{\psi^-} | \mbox{2 photons, interfering, in} ) =  \nonumber \\
\hspace{3mm} C \times P(\ket{\psi^-} | \mbox{2 photons, 1 spatial mode, out}) + \nonumber \\
\hspace{3mm} D \times P(\ket{\psi^-} | \mbox{2 photons, 2 spatial modes, 2 temporal modes, out}).
\label{eqn:2_photons_interfering}
\end{eqnarray}
\noindent

\subsection{Aggregate probability for projections onto $\ket{\psi^-}$ \label{sec:visibility}}
Now that we have calculated the conditional probabilities of a detection pattern indicating $\ket{\psi^-}$ for various inputs to the beam splitter, let us consider with what probability each case occurs. This requires that we know the photon number distribution of the pulses arriving at Charlie's beam splitter from Alice and Bob, which can be computed based on the photon number distribution at the sources and the properties of the quantum channels.    For the following discussion, we assume that the channels from Alice to Charlie, and from Bob to Charlie are characterized by the loss $t_A$ and $t_B$, respectively, yielding pulses with number distribution $\mathbb{D}$ and mean photon number, $\mu t_A$ and $\sigma t_B$, respectively.  This is equivalent to assuming that no PNS attack takes place, which was ensured by performing experiments with the entire setup (including the fiber transmission lines) inside a single laboratory in which no eavesdropping took place during the experiments. We limit our discussion to the cases with two or less photons at the input of the beam splitter (but recall that the actual calculation includes up to three photons).   Hence, the cases we consider and their probabilities of occurrence, $P_O$, are given by:

\begin{itemize}
\item 0 photons at the input  from both sources:  $P_O=\mathbb{D}_0(\mu t_A)\mathbb{D}_0(\sigma t_B)$
\item 1 photon at the input from Alice and 0  photons from Bob:  $P_O=\mathbb{D}_1(\mu t_A)\mathbb{D}_0(\sigma t_B)$
\item 0 photons at the input  from Alice and 1 photon from Bob:  $P_O=\mathbb{D}_0(\mu t_A)\mathbb{D}_1(\sigma t_B)$
\item 2 photons at the input  from Alice and 0 photons from Bob:  $P_O=\mathbb{D}_2(\mu t_A)\mathbb{D}_0(\sigma t_B)$
\item 0 photons at the input from Alice and 2 photons from Bob:  $P_O=\mathbb{D}_0(\mu t_A)\mathbb{D}_2(\sigma t_B)$
\item 1 photon at the input from both sources:  $P_O=\mathbb{D}_1(\mu t_A)\mathbb{D}_1(\sigma t_B)$
\end{itemize}

\noindent
where we denote the probability of having $i$ photons from a  distribution $\mathbb{D}$ with mean number $\mu$ as $\mathbb{D}_i(\mu)$.  For each of these cases, we have already computed the probability that Charlie obtains the detection pattern associated with the $\ket{\psi^-}$-state for arbitrary input states  of the photons (as defined in Eq.~(\ref{eqn:photon_state})).  When zero or one photons arrive at the beam splitter, Eq.~(\ref{eqn:0_photons}) and Eq.~(\ref{eqn:1_photon}) are used, respectively.  In the case in which two photons arrive  from the same source, Eq.~(\ref{eqn:2_photons_non_interfering}) is used.  Finally, in the case in which one photon arrives from each source at the beam splitter, Eq.~(\ref{eqn:2_photons_interfering}) would be used in the ideal case.  However, perfect indistinguishability of the photons cannot be guaranteed in practice. We characterize the degree of indistinguishability by the visibility, $V$, that we would observe in a closely-related Hong-Ou-Mandel (HOM) interference experiment~\cite{Hong1987} with single-photon inputs. Taking into account partial distinguishability, the probability of finding a detection pattern corresponding to the projection onto $\ket{\psi^-}$ is given by

\begin{eqnarray}
P(\ket{\psi^-} | \mbox{2 photons, visibility $V$, in}) = \nonumber \\
\hspace{3mm} VP(\ket{\psi^-} | \mbox{2 photons, interfering, in}) \nonumber \\
\hspace{3mm} + (1-V)P(\ket{\psi^-} | \mbox{2 photons, non-interfering, in}).
\label{eqn:2_photons_V}
\end{eqnarray}

\noindent Equations \ref{eqn:0_photons}-\ref{eqn:2_photons_V} detail all possible causes for observing the detection pattern associated with a projection onto the $\ket{\psi^-}$ Bell state, if up to two photons at the beam splitter input are taken into account. We remind the reader that all calculations in the following sections take up to three photons at the input of the beam splitter into account. To calculate the gains, $Q_{\mu\sigma}^{x,z}$, using these equations, we need only substitute in the correct values of $\mu$, $\sigma$, $t_A$, $t_B$, $m^{x,z}$, $b^{x,z}$, and $\Delta\phi^{x,z}$ for the cases in which Alice and Bob both sent attenuated light pulses in the x-basis or z-basis, respectively. The error rates, $e_\mu^{x,z}$, can then be computed by separating the projections onto $\ket{\psi^-}$ into those where Alice and Bob sent photons in different states (yielding correct key bits) and in the same state (yielding erroneous key bits). More precisely, the error rates, $e_{\mu\sigma}^{x,z}$, are calculated as $e_{\mu\sigma}^{x,z}=p^{x,z}_{wrong}/(p^{x,z}_{correct}+p^{x,z}_{wrong}$) where $p^{x,z}_{wrong}$ ($p^{x,z}_{correct}$) denotes the probability for detections yielding an erroneous (correct) bit in the $x$ (or $z$)-key.

\section{Characterizing experimental imperfections\label{sec:characterization}}
The parameters used to model our system are derived from data established through independent measurements. To test our model, the characterization of experimental imperfections in our MDI-QKD implementation~\cite{Rubenok2013} is very technical at times. It can be broken down into time-resolved energy measurements at the single photon level (required to extract $\mu$, $\sigma$, $b^{x,z}$ and $m^{x,z}$ for Alice and Bob, as well as dark count and afterpulsing probabilities), measurements of phase (required to establish $\phi^{x,z}$ for Alice and Bob), and visibility measurements. In the following paragraphs we describe the procedures we followed to obtain these parameters from our system. 

\subsection{Our MDI-QKD implementation}
In our implementation of MDI-QKD~\cite{Rubenok2013} Alice's and Bob's setups are identical. Each setup consists of a CW laser with large coherence time, emitting at 1550nm wavelength. Time-bin qubits, encoded into single photon-level light pulses with Poissonian photon number statistics, are created through an attenuator, an intensity modulator and a phase modulator located in a temperature controlled box. More precisely, the intensity modulator is used to tailor pulse pairs out of the cw laser light, the phase modulator is used to change their relative phase, and the attenuator attenuates these pulses to the single-photon level. The two temporal modes defining each time-bin qubit are of 500 ps (FWHM) duration and are separated by 1.4 ns. Each source generates qubits at 2 MHz rate.

We emphasize that our qubit generation procedure justifies the assumption of a pure state in Eq.~(\ref{eqn:photon_state}). Indeed, all photons, including background photons due to light leaking through imperfect intensity modulators, have to be generated by the CW lasers whose coherence times exceeds the separation between the temporal modes $\ket{0}$ and $\ket{1}$~\cite{note4}. Note that in all experiments reported to date  \cite{Rubenok2013,Liu2012, daSilva2012, Tang2013} background photons always add coherently to the modes describing qubits, making our pure-state description widely applicable.

\begin{figure}[h!]
\centering \includegraphics[width=13cm]{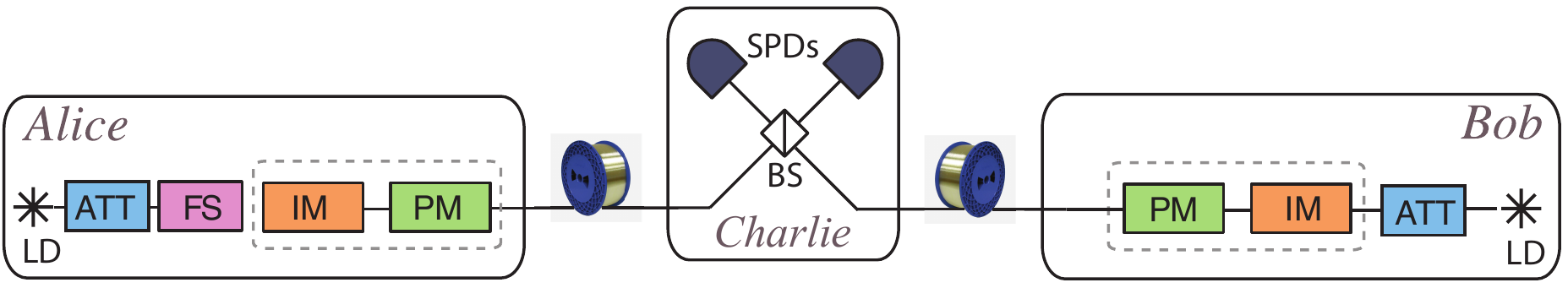}
	\begin{center}
		\caption{\label{fig:setup} Time-bin qubits are created at Alice's and Bob's through a CW laser (LD), attenuator (ATT), and frequency shifter (FS) and temperature-controlled intensity (IM) and phase (PM) modulator. The projective measurements are done at Charlie's via a beam splitter (BS) and two single photon detectors (SPDs).}
	\end{center}
\end{figure}

The time-bin qubits are sent to Charlie through an optical fiber link. The link consisted of spooled fiber (for the measurements in which Alice, Bob and Charlie were all located in the same laboratory) or deployed fiber (for the measurements in which the three parties were located in different locations within the city of Calgary). We remind the reader that all pulses arriving at Charlie's are phase randomized, due to the use of long fibers. Charlie performs a BSM on the qubits he receives using a 50/50 beamsplitter and two SPDs. See Figure~\ref{fig:setup}. Note that, in order to perform a Bell state measurement the photons arriving to Charlie must be indistinguishable in all degrees of freedom: polarization, frequency, time and spatial mode. The indistinguishability of the photons is assessed through a Hong-Ou-Mandel interference measurement~\cite{Hong1987}. As our system employs attenuated laser pulses, the maximum visibility we can obtain in this measurement is $V_{max}=50\%$ (and not 100\% as it would be with single photons) \cite{Mandel83}. In our implementation the visibility measurements resulted in $V=(47 \pm 1)$, irrespective of whether they were taken with spooled fiber inside the lab, or over deployed fiber. 

\subsection{Time-resolved energy measurements \label{sec:time}}
First, we characterize the dark count probability per time-bin, $P_{d}$, of the SPDs (InGaAs-avalanche photodiodes operated in gated Geiger mode~\cite{Stucki2001}) by observing their count rates when the optical inputs are disconnected.  We then send attenuated laser pulses so that they arrive just after the end of the 10 ns long gate that temporarily enables single photon detection.  The observed change in the count rate is due to background light transmitted by the intensity modulators (whose extinction ratios are limited) and allows us to establish $b^{x,z}$ (per time-bin) for Alice and Bob. Next, we characterize the afterpulsing probability per time-bin, $P_{a}$, by placing the pulses within the gate, and observing the change in count rate in the region of the gate prior to the arrival of the pulse.  The afterpulsing model we use to assess $P_{a}$  from these measurements is described below.

\begin{figure}[t]
\centering \includegraphics{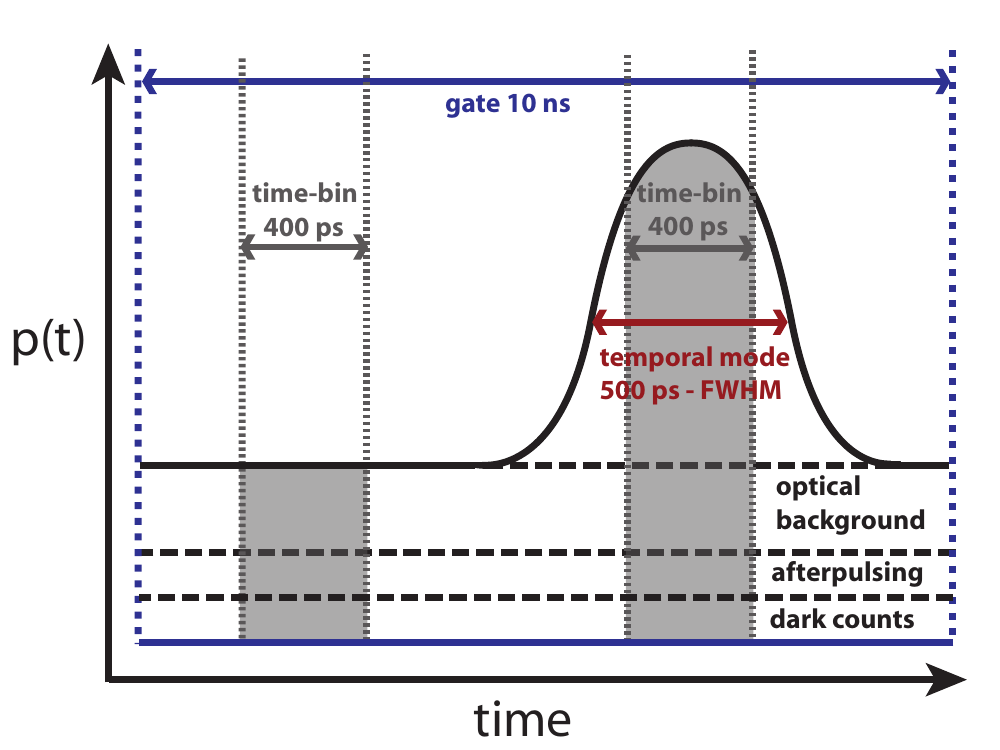}
	\begin{center}
		\caption{\label{fig:time-bin} Sketch (not to scale) of the probability density p(t) for a detection event to occur as a function of time within one gate. Detection events can arise from a photon within an optical pulse (depicted here as a pulse in the late temporal mode), or be due to optical background, a dark count, or afterpulsing. Also shown are the 400 ps wide time-bins. Within the early time-bin only optical background, dark counts and afterpulsing give rise to detection events in this case. Note that the width of the temporal mode exceeds the widths of the time-bins.}
	\end{center}
\end{figure}

Once the background light and the sources of detector noise  are characterized, the values of $m^{x,z}$  can be calculated by generating all required states and observing the count rates in the two time-bins corresponding to detecting photons generated in early and late temporal modes. Observe that $m^{z=1}$ for photons generated in state $\ket{1}$ (the late temporal mode) is zero, since all counts in the early time-bin are attributed to one of the three sources of background described above. Furthermore, we observed that $m^{z=0}$ for photons generated in the $\ket{0}$ state (the early temporal mode) is smaller than one due to electrical ringing in the signals driving the intensity modulators. Note that, in our implementation, the duration of a temporal mode exceeds the width of a time-bin, i.e. it is possible to detect photons outside a time-bin (see Figure~\ref{fig:time-bin} for a schematical representation). Hence, it will be useful to also define the probability for detecting a photon arriving at any time during a detector gate; we will refer to this quantity as $\eta_{gate}$.The count rate per gate, after having subtracted the rates due to background and detector noise, together with the detection efficiency, $\eta_{gate}$ ($\eta_{gate}$, as well as $\eta$, have been characterized previously based on the usual procedure~\cite{Stucki2001}), allows calculating the mean number of photons per pulse from Alice or Bob ($\mu$ or $\sigma$, respectively). The efficiency coefficient relevant for our model, $\eta$, is smaller than $\eta_{gate}$. Finally, we point out that the entire characterization described above was repeated for all experimental configurations investigated (the configurations are detailed in Table~\ref{tab:results}). We found all parameters to be constant in $\mu\sigma t_At_B$, with the obvious exception of the afterpulsing probability.

\subsection{Phase measurements \label{sec:phase}}
To detail the assessment of the phase values $\phi^{x,z}$ determining the superposition of photons in early and late temporal modes, let us assume for the moment that the lasers at Alice's and Bob's emit light at the same frequency. First, we defined the phase of Bob's $\ket{+}$ state to be zero (this can always be done by appropriately defining the time difference between the two temporal modes $\ket{0}$ and $\ket{1}$). Next, to measure the phase describing any other state (generated by either Alice or Bob) with respect to Bob's $\ket{+}$ state, we sequentially send unattenuated laser pulses encoding the two states through a common reference interferometer. This reference interferometer featured a path-length difference equal to the time-difference between the two temporal modes defining AliceÕs and Bob's qubits.  For the phase measurement of qubit states $\ket{+}$ and $\ket{-}$ (generate by Alice), and $\ket{-}$ generated by Bob), first, the phase of the interferometer was set such that Bob's $\ket{+}$ state generated equal intensities in each output of the interferometer (i.e. the interferometerÕs phase was set to $\pi/4$). Thus, sending any of the other three states  through the interferometer and comparing the output intensities, we can calculate the phase difference. We note that any frequency difference between Alice's and Bob's lasers results in an additional phase difference. Its upper bound for our maximum frequency difference of 10 MHz is denoted by $\phi_{freq}$.

\subsection{Measurements of afterpulsing \label{sec:ap}}
We now turn to the characterization of afterpulsing.  After a detector click (or detection event, which includes photon detection, dark counts and afterpulsing), the probability of an afterpulse occuring due to that detection event decays exponentially with time.  The SPDs are gated, with the afterpulse probability per gate being a discrete sampling of the exponential decay.  This can be expressed using a geometric distribution: supposing a detection event occurred at gate $k=-1$, the probability of an afterpulse occuring in gate $k$ is given by $P_k =\alpha p(1-p)^k$.  Thus, if there are no other sources of detection events, the probability of an afterpulse occuring due to a detection event is given by $\sum_{k=0}^\infty \alpha p(1-p)^k$.

\begin{figure}[b]
 	\centering \includegraphics[width=8cm]{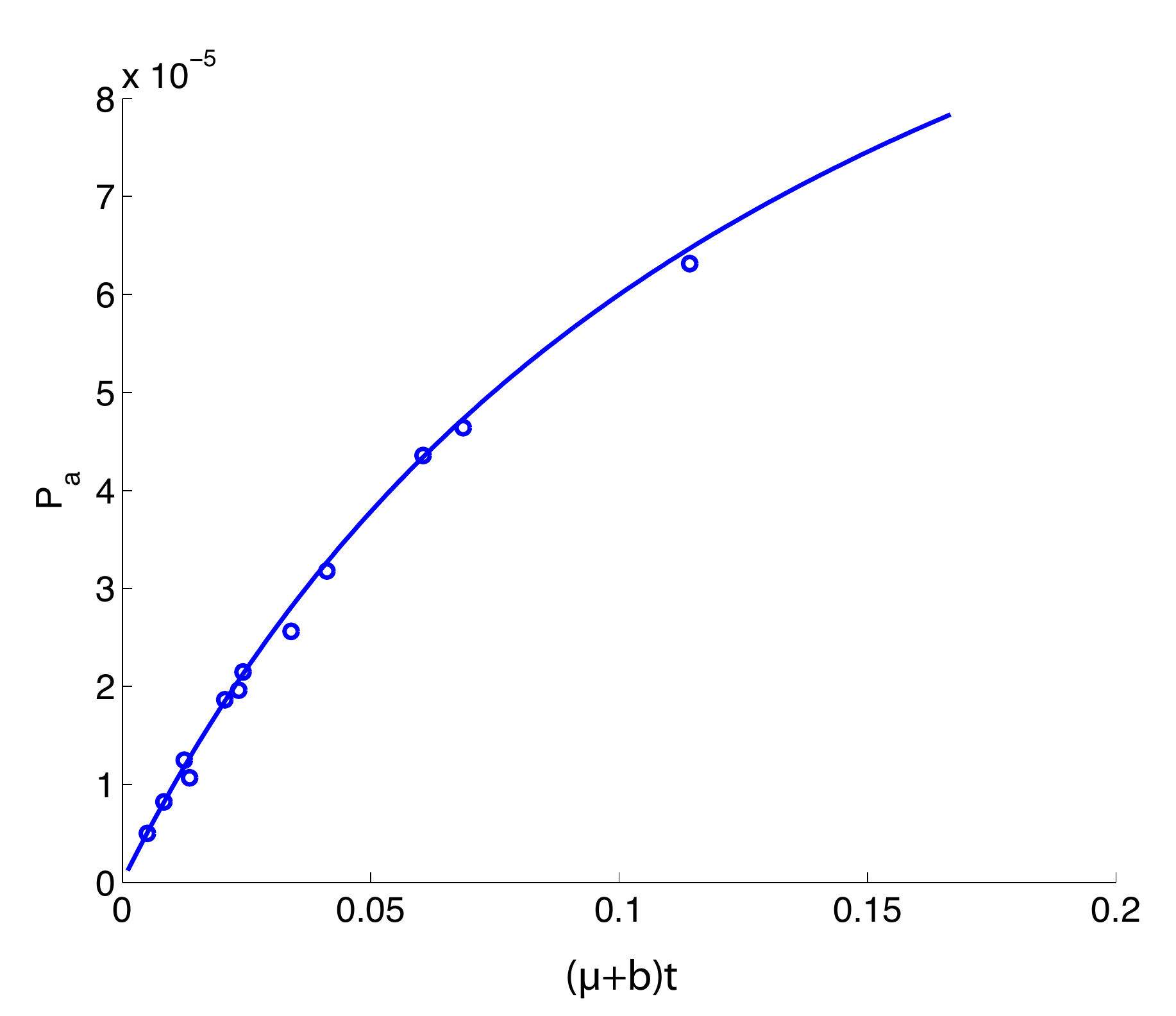}
 \begin{center}
 \caption{\label{fig:ap}Afterpulse probability per time-bin as a function of the average number of photons arriving at the detector per gate.}
 \end{center}
\end{figure}

In a realistic situation, the geometric distribution for the afterpulses will be cut off by other detection events, either stemming from photons, or dark counts.  In addition, the SPDs have a deadtime after each detection event during which the detector is not gated until $k \ge k_{dead}$ (note that time and the number of gates applied to the detector are proportional).  The deadtime can simply be accounted for  by starting the above summation at $k=k_{dead}$ rather than $k=0$.  However, for an afterpulse to occur  during the $k^{{th}}$ gate following a particular detection event, no other detection events must have occured in prior gates.  This leads to the following equation for the probability of an afterpulse per detection event:

\begin{equation}
P(\mbox{a,det}) = \sum_{k=k_{dead}}^\infty \left (\gamma \times \upsilon \times \rho \times P_k\right)
\label{eqn:ap_per_det}
\end{equation}

\noindent
where: 
\begin{eqnarray}
\hspace{3mm}\gamma = \left(1-\mu_\mathrm{avg}(\mu,\sigma,t_A,t_B)\eta_{gate}\right)^{k-k_{dead}}
\nonumber \\
\nonumber \\
\hspace{3mm} \upsilon  = (1 - P_{d,gate})^{k-k_{dead}}
\nonumber \\
\nonumber \\
\hspace{3mm} \rho = \prod_{j=k_{dead}}^{k-1}1 - \alpha p(1-p)^{j} \nonumber \\
\nonumber \\
\hspace{3mm} P_k = \alpha p(1-p)^k
\end{eqnarray}

\noindent
and $P_{d,gate}$ denotes the detector dark count probability per gate (as opposed to per time-bin), and $\mu_\mathrm{avg}(\mu,\sigma,t_A,t_B)$ expresses the average number of photons present on the detector during each gate as follows:

\begin{equation}
\mu_\mathrm{avg}(\mu,\sigma,t_A,t_B) = \frac{(\mu+b_A)t_A + (\sigma + b_B)t_B}{2},
\label{eqn:mu_avg}
\end{equation}
\noindent
where $b_A$ and $b_B$ characterize the amount of background light per gate from Alice and Bob, respectively, and the factor of $\frac{1}{2}$ comes from Charlie's beam splitter. The terms in the sum of Eq.~(\ref{eqn:ap_per_det}) describe the probabilities of neither having an optical detection ($\gamma$), either caused by a modulated pulse or background light, nor a detector dark count ($\upsilon$) in any gate before and including gate $k$,  and not having an afterpulse in any gate before gate $k$  ($\rho$), followed by an afterpulse in gate $k$ ($P_k$).  Equation~(\ref{eqn:ap_per_det}) takes into account that  afterpulsing within each time-bin is influenced by all detections within each detector gate, and not only those happening within the time-bins that we post-select when acquiring experimental data.

The afterpulse probability, $P_{a,gate}$, for given $\mu$, $\sigma$, $t_A$ and $t_B$ can then be found by multiplying Eq.~(\ref{eqn:ap_per_det}) by the total count rate

\begin{equation}
P_{a,gate} = \left(\mu_\mathrm{avg}(\mu,\sigma,t_A,t_B)\eta_{gate} + P_{d,gate} + P_{a,gate}\right)P(\mbox{a,det}).
\label{eqn:ap_per_gate_wip}
\end{equation}
\noindent
This equation expresses  that afterpulsing can arise from prior afterpulsing, which explains the appearance of $P_{a,gate}$ on both sides of the equation. Equation~(\ref{eqn:ap_per_gate_wip}) simplifies to

\begin{equation}
P_{a,gate} = \frac{\left (\mu_\mathrm{avg}(\mu,\sigma,t_A,t_B)\eta_{gate} + P_{d,gate}\right )P(\mbox{a,det})}{1-P(\mbox{a,det})}.
\label{eqn:ap_per_gate}
\end{equation}
\noindent
Finally, to extract the afterpulsing probability per time-bin, $P_{a}(\mu,\sigma,t_A,t_B)$, we note that we found that the distribution of afterpulsing across the gate to be the same as the distribution of dark counts across the gate.  Hence,
\begin{equation}
P_{a}(\mu,\sigma, t_A,t_B)= P_{a,gate}\frac{P_{d}}{P_{d,gate}}.
\end{equation}
\noindent
Fitting our afterpulse model to the measured afterpulse probabilities, we find $\alpha=1.79 \times 10^{-1}$, $p= 2.90 \times 10^{-2}$, and $\frac{P_{d}}{P_{d,gate}} =4.97 \times 10^{-2}$ for $k_{dead} = 20$.  The fit, along with the measured values, is shown in Figure~\ref{fig:ap} as a function of the average number of photons arriving at the detector per gate $\mu_\mathrm{avg}(\mu,\sigma,t_A,t_B)$. 

\noindent A summary of all the values obtained through these measurements is shown in Table~\ref{tab:params}.

\begin{table}[h!]
\caption{\label{tab:params}Experimentally established values for all parameters required to describe the generated quantum states, as defined in Eq.~(\ref{eqn:photon_state}), as well as two-photon interference parameters and detector properties.}
\footnotesize\rm
\begin{tabular*}{\textwidth}{@{}l*{15}{@{\extracolsep{0pt plus12pt}}l}}
%\br
\hline
 Parameter&Alice's value&Bob's value\\
\hline
$b^{z=0}=b^{z=1}$ & $(7.12 \pm 0.98) \times 10^{-3}$ &  $(1.14 \pm 0.49) \times 10^{-3}$ \\
$b^{x=-}=b^{x=+}$ & $(5.45 \pm 0.37) \times 10^{-3}$ &  $(1.14 \pm 0.49) \times 10^{-3}$ \\
$m^{z=0}$ & $0.9944 \pm 0.0018$ & $0.9967 \pm 0.0008$ \\
$m^{z=1}$ & 0 & 0 \\
$m^{x=+}=m^{x=-}$ & $0.4972 \pm 0.011 $ & $0.5018 \pm 0.0080$ \\
$\phi^{z=0}=\phi^{z=1}=\phi^{x=+}$ [rad] & 0 & 0 \\
$\phi^{x=-}$ [rad] & $\pi + (0.075 \pm 0.015)$ & $\pi - (0.075 \pm 0.015)$ \\
%\br
\hline
 Parameter&Value\\
%\mr
\hline
$\left|\phi_{freq}\right|$ [rad]& $< 0.088$ \\
$V$ & $0.94 \pm 0.02 $\\
$P_{d}$ & $(1.83 \pm 0.77) \times 10^{-5}$ \\
$\eta_{gate}$ & 0.2 \\
$\eta$ & 0.145 \\
%\br
\hline
\end{tabular*}
\end{table}

\section{Testing the model, and real-world tests} \label{sec:verification}

\subsection{Comparing modelled with actual performance}
\begin{figure}[h!]
     \begin{center}
            \includegraphics[width=13cm]{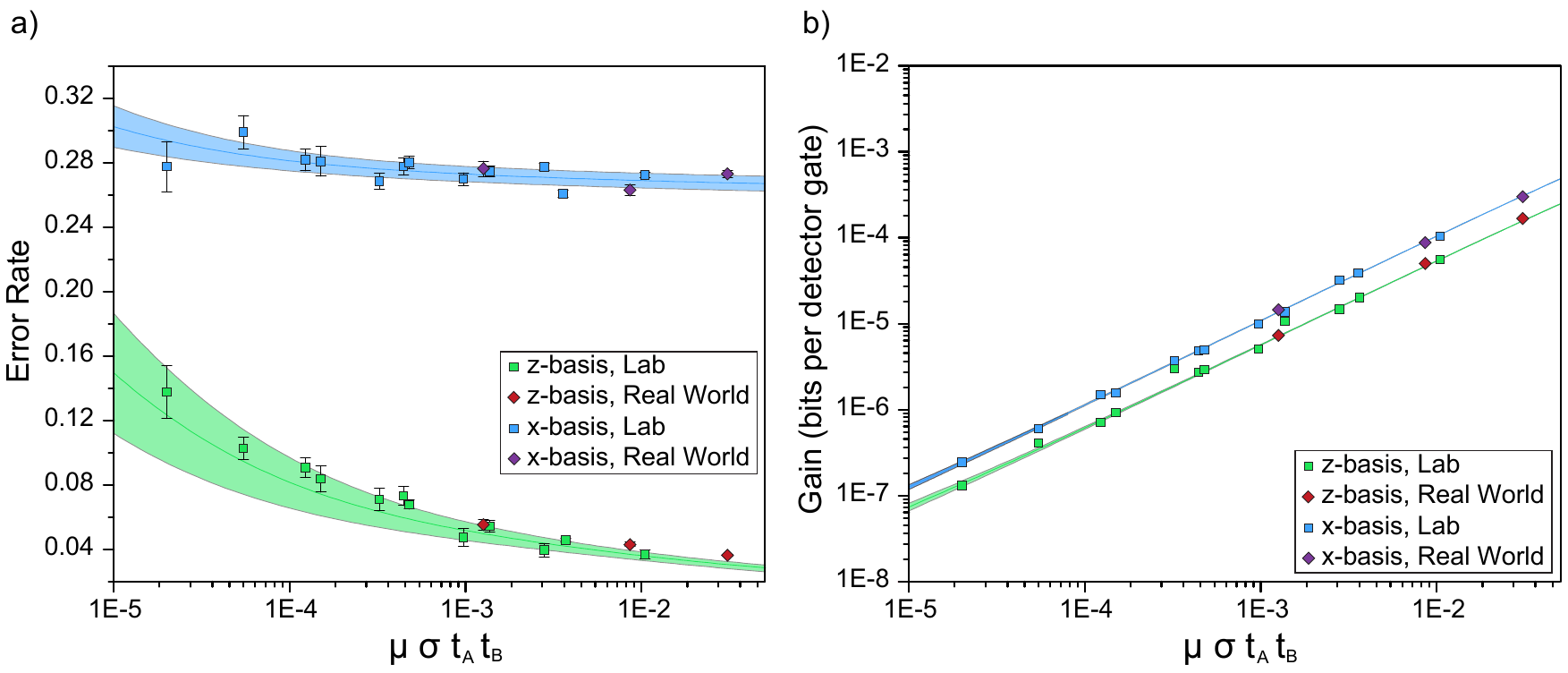}
    \caption{\label{fig:results} Modelled and measured results. Figure a) shows the plot for the error rates in the $z$-basis (green band) and in the $x$-basis (blue band) as a function of the mean photon number per pulse sent by Alice ($\mu$) and Bob ($\sigma$) multiplied by the channel transmissions ($t_A$ and $t_B$).  Figure b) shows the plot of the gains  as a function of $\mu \sigma t_A t_B $. The $z$-basis is shown in green and the $x$-basis is shown in blue. For both figures the results of the measurements done in the laboratory are shown with squares (blue or green) and the measurements done over deployed fiber are shown with diamond (red and purple). The difference in gains and error rates in the x- and the z-basis, respectively is due to the fact that, in the case in which one party sends a laser pulse containing more than one photon and the other party sends zero photons, projections onto the $\ket{\psi^-}$ Bell state can only occur if both pulses encode qubits belonging to the x-basis. The Bell state projection cannot occur if both prepare qubits belonging to the z-basis (we ignore detector noise for the sake of this argument). This causes increased gain for the x-basis and, due to  an error rate of 50\% associated with these projections, also an increased error rate for the x-basis.}
        \end{center}
\end{figure}

To test our model, and to verify our ability to perform, in principle, QKD with deployed (real-world) fiber, we now compare the model's predictions with experimental data obtained using the QKD system characterized by the parameters listed in Table~\ref{tab:params}.  We performed experiments in two configurations: inside the laboratory using spooled fiber (for four different distances between Alice and Bob ranging between 42 km and 103 km), and over deployed fiber (18 km). The first configuration allows testing the model, and the second configurations shines light on our system's capability to compensate for environment-induced perturbations, e.g. due to temperature fluctuations. For each test, three different mean photon numbers (0.1, 0.25 and 0.5) were used. All the configurations tested (as well as the specific parameters used in each test) and the results obtained are listed in Table~\ref{tab:results}.  In Figure~\ref{fig:results} we show the simulated values for the error rates ($e^{z,x}$) and gains ($Q^{z,x}$) predicted by the model as a function of $\mu\sigma t_A t_B$. The plot includes uncertainties from the measured parameters, leading to a range of values (bands) as opposed to single values.  The figure also shows the experimental values of $e^{z,x}$ and $Q^{z,x}$ from our MDI-QKD system in both the laboratory environment and over deployed fiber.

Considering the data taken inside the lab, the modelled values and the experimental results agree within experimental uncertainties over three orders of magnitude. This shows that the model is suitable for predicting error rates and gains. In turn, this allows us to optimize performance of our QKD systems in terms of secret key rate (see section \ref{sec:optimization}). In particular, the model allows optimizing the mean photon number per pulse that Alice and Bob use to encode signal and decoy states as a function of transmission loss, and identifying rate-limiting components. 

Furthermore, the measurement results over deployed fibre are also well described by the same model, indicating that this more-difficult measurement worked correctly.  The increased difficult across real-world fiber arises due to the fact that BSMs require incoming photons to be indistinguishable in all degrees of freedom (i.e. arrive within their respective coherence times, with identical polarization, and with large spectral overlap).  As we have shown in~\cite{Rubenok2013}, time-varying properties of optical fibers in the outside environment (e.g. temperature dependent polarization and travel-time changes) can remove indistinguishability in less than a minute.  Active stabilization of these properties is thus required to achieve functioning BSMs and, in fact, three such stabilization systems were deployed during the MDI-QKD measurements presented here (more details are contained in~\cite{Rubenok2013}). That our measurement results agree with the predicted values of the model demonstrates that the impact of environmental perturbations on the ability to perform Bell state measurements is negligible (which is the same conclusion drawn in~\cite{Rubenok2013}).

\begin{table}[t]
\caption{\label{tab:results} Measured error rates, $e_{\mu\sigma}^{x,z}$, and  gains, $Q_{\mu\sigma}^{x,z}$, for different mean photon numbers, $\mu$ and $\sigma$ (where $\mu=\sigma$), lengths of fiber connecting Alice and Charlie, and Charlie and Bob, $\ell_{A}$ and $\ell_{B}$, respectively, and total transmission loss, $l$. The last set of data details real-world measurements using deployed fiber.  Uncertainties are calculated using Poissonian detection statistics.}
\footnotesize\rm
\begin{tabular*}{\textwidth}{@{}l*{15}{@{\extracolsep{0pt plus12pt}}l}}
%\begin{tabular*}
%\br
\hline
Fiber & $\mu = \sigma$ & $\ell_{A}$ & $\ell_{B}$ & total loss  & $Q_{\mu\sigma}^x$ & $Q_{\mu\sigma}^z$ & $e_{\mu\sigma}^x$ & $e_{\mu\sigma}^z$\\
   ~ & ~ &  [km] & [km] &  $l$ [dB] & ~ & ~ & ~ & ~\\
%  \mr
\hline
             & $0.49(2)$ &   &  &  & $1.045(4) \times 10^{-4}$ & $5.57(8) \times 10^{-5}$ & 0.272(2) & 0.037(3) \\
  Spool & $0.254(9)$ & 30.98 & 11.75 & 13.6& $3.20(2) \times 10^{-5}$ & $1.47(3) \times 10^{-5}$ & 0.277(2) & 0.040(4) \\
             & $0.101(4)$ &   &   &  & $4.84(6) \times 10^{-6}$ & $2.72(6) \times 10^{-6}$ & 0.278(5) & 0.073(6) \\
%   \mr
\hline
              & $0.49(2) $ &   &  &  &  $3.92(2) \times 10^{-5}$ & $2.02(1) \times 10^{-5}$ & 0.261(2) & 0.046(1) \\
  Spool & $0.25(1) $ & 40.80 & 40.77 & 18.2& $9.87(9) \times 10^{-6}$ & $5.1(1) \times 10^{-6}$ & 0.270(4) & 0.047(5) \\
              & $0.099(4)$ &   &   &  & $1.57(3) \times 10^{-6}$ & $9.2(3) \times 10^{-7}$ & 0.281(9) & 0.084(8) \\
 %    \mr
 \hline
              & $0.50(2)$ &   &  &  &  $1.37(1) \times 10^{-5}$ & $1.07(2) \times 10^{-5}$ & 0.275(3) & 0.054(4) \\
  Spool & $0.24(1)$ &51.43 & 32.19 & 22.7& $3.73(4) \times 10^{-6}$ & $3.01(8) \times 10^{-6}$ & 0.269(5) & 0.071(7) \\
             & $0.100(6)$ &   &   &  & $6.0(1) \times 10^{-7}$ & $4.07(9) \times 10^{-7}$ & 0.30(1) & 0.103(7) \\
  % \mr
  \hline
              & $0.50(5) $ &   &  &  &  $4.96(4) \times 10^{-6}$ & $2.94(3) \times 10^{-6}$ & 0.280(4) & 0.068(3) \\
  Spool & $0.25(1) $ & 61.15 & 42.80 & 27.2& $1.50(2) \times 10^{-6}$ & $7.1(2) \times 10^{-7}$ & 0.282(7) & 0.091(6) \\
              & $0.103(5)$ &   &   &  & $2.45(9) \times 10^{-7}$ & $1.31(6) \times 10^{-7}$ & 0.28(2) & 0.14(2) \\
 %  \mr
 \hline
                     & $0.50(2) $ &   &  &  &  $3.01(1) \times 10^{-4}$ & $1.667(8) \times 10^{-4}$ & 0.273(2) & 0.0362(7) \\
  Deployed & $0.26(1) $ & 12.4 & 6.2 & 9.0& $8.78(6) \times 10^{-5}$ & $5.01(4) \times 10^{-5}$ & 0.263(3) & 0.043(1) \\
                     & $0.100(4) $ &   &   &  & $1.45(2) \times 10^{-5}$ & $7.3(1) \times 10^{-7}$ & 0.276(5) & 0.055(3) \\
  % \br
  \hline
\end{tabular*}
\end{table}

\section{Optimization of system performance}\label{sec:optimization}

\subsection{Decoy-state analysis}\label{sec:decoy}
To calculate secret key rates for various system parameters, which allows optimizing these parameters, first, it is necessary to compute the gain, $Q_{11}^{z}$, and the error rate, $e_{11}^{x}$, that stem from events in which both sources emit a single photon.  
We consider the three-intensity decoy state method for the MDI-QKD protocol proposed in~\cite{Wang2013}, which derives a lower bound for the secret key rate using lower bounds for $Q_{11}^{x,z}$ and an upper bound for $e_{11}^x$. Note that we assume here that the the only effect of imperfectly generated qubit states on the secret key rate that we consider here is that it increases the error rates (further considerations require advancements to security proofs, which are under way~\cite{Wang2013,Tamaki11}) increases of error rates.  

We denote the signal, decoy, and vacuum intensities by $\mu_s$, $\mu_d,$ and $\mu_v$, respectively, for Alice, and, similarly, as $\sigma_s$, $\sigma_d$, and $\sigma_v$ for Bob. Note that $\mu_v = \sigma_v = 0$ by definition. This decoy analysis assumes that perfect vacuum intensities are achievable, which may not be correct in an experimental implementation. However, note that, first, intensity modulators with more than 50 dB extinction ratio exist, which allows obtaining almost zero vacuum intensity, and second, that a similar decoy state analysis with non-zero vacuum intensity values is possible as well~\cite{Xu2013}.  
For the purpose of this analysis, we take both channels to have the same transmission coefficients (that is $t_A = t_B \equiv t$), according to our experimental configuration, and  Alice and Bob hence both select the same mean photon numbers for each of the three intensities (that is $\mu_s = \sigma_s \equiv \tau_s$, $\mu_d = \sigma_d \equiv \tau_d$, and $\mu_v = \sigma_v \equiv \tau_v$).  Additionally, for compactness of notation, we omit the $\mu$ and $\sigma$ when describing the gains and error rates (e.g. we write $Q_{ss}^z$ to denote the gain in the z-basis when Alice and Bob both send photons using the signal intensity).  Under these assumptions, the lower bound on $Q_{11}^{x,z}$ is given by

\begin{equation}
Q_{11}^{x,z} \ge \frac{ \mathbb{D}_1(\tau_s)\mathbb{D}_2(\tau_s)\big(Q_{dd}^{x,z} - Q_0^{x,z}(\tau_d)\big) - \mathbb{D}_1(\tau_d)\mathbb{D}_2(\tau_d)\big(Q_{ss}^{x,z} - Q_0^{x,z}(\tau_s)\big)  }{\mathbb{D}_1(\tau_s)\mathbb{D}_1(\tau_d)\big(\mathbb{D}_1(\tau_d)\mathbb{D}_2(\tau_s)-\mathbb{D}_1(\tau_s)\mathbb{D}_2(\tau_d)\big)},
\label{eqn:Q11_bound}
\end{equation}

\noindent
where the various $\mathbb{D}_i(\tau)$ denote the probability that a pulse with photon number distribution $\mathbb{D}$ and mean $\tau$ contains exactly $i$ photons, and $Q_{0}^{x,z}(\tau_d)$ and $Q_{0}^{x,z}(\tau_s)$ are given by

\begin{eqnarray}
Q_{0}^{x,z}(\tau_d) & = & \mathbb{D}_0(\tau_d)Q_{vd}^{x,z} + \mathbb{D}_0(\tau_d)Q_{dv}^{x,z} - \mathbb{D}_0(\tau_d)^2Q_{vv}^{x,z},\\
Q_{0}^{x,z}(\tau_s) & = & \mathbb{D}_0(\tau_s)Q_{vs}^{x,z} + \mathbb{D}_0(\tau_s)Q_{sv}^{x,z} - \mathbb{D}_0(\tau_s)^2Q_{vv}^{x,z}.
\label{eqn:Q0}
\end{eqnarray}

\noindent
The error rate $e_{11}^x$ can then be computed as

\begin{equation}
e_{11}^x \le \frac{e_{dd}^xQ_{dd}^x - \mathbb{D}_0(\tau_d)e_{vd}^xQ_{vd}^x - \mathbb{D}_0(\tau_d)e_{dv}^xQ_{dv}^x + \mathbb{D}_0(\tau_d)^2e_{vv}^xQ_{vv}^x}{\mathbb{D}_1(\tau_d)^2Q_{11}^x},
\label{eqn:e11_bound}
\end{equation}

\noindent
where the upper bound holds if a lower bound is used for $Q_{11}^x$. Note that $Q_{11}^{x,z}$, $Q_{0}^{x,z}(\tau_d)$, $Q_{0}^{x,z}(\tau_s)$ and $e_{11}^x$ (Eqs.~(\ref{eqn:Q11_bound}-\ref{eqn:e11_bound})) are uniquely determined through measurable gains and error rates.

\subsection{Optimization of signal and decoy intensities}
For each set of experimental parameters (i.e. distribution function $\mathbb{D}$, channel transmissions and all parameters describing imperfect state preparation and measurement), the secret key rate (Eq.~(\ref{eq:secret_key_rate})) can be maximized by properly selecting the intensities of the signal and decoy states ($\tau_s$ and $\tau_d$, respectively). Here we consider its optimization as a function of the total transmission (or distance) between Alice and Bob. We make the assumptions that both the channel between Alice and Charlie and the channel between Bob and Charlie have the same transmission coefficient, $t$, and that Alice and Bob use the same signal and decoy intensities. We considered values of $\tau_d$ in the range $0.01 \le \tau_d < 0.99$ and values of $\tau_s$ in the range $\tau_d < \tau_s \le 1$.  An exhaustive search computing the secret key rate for an error correction efficiency $f=1.14$~\cite{Sasaki2011} is performed from 2~km to 200~km total distance (assuming 0.2 dB/km loss), with increments of 0.01 photons per pulse for both $\tau_s$ and $\tau_d$.  For each point, the model described in section~\ref{sec:model} is used to compute all the experimentally accessible quantities required to compute secret key rates using the three-intensity decoy state method summarized in Eqs.~(\ref{eqn:Q11_bound}-\ref{eqn:e11_bound}).

\begin{figure}[t]
\centering  \includegraphics[width=13cm]{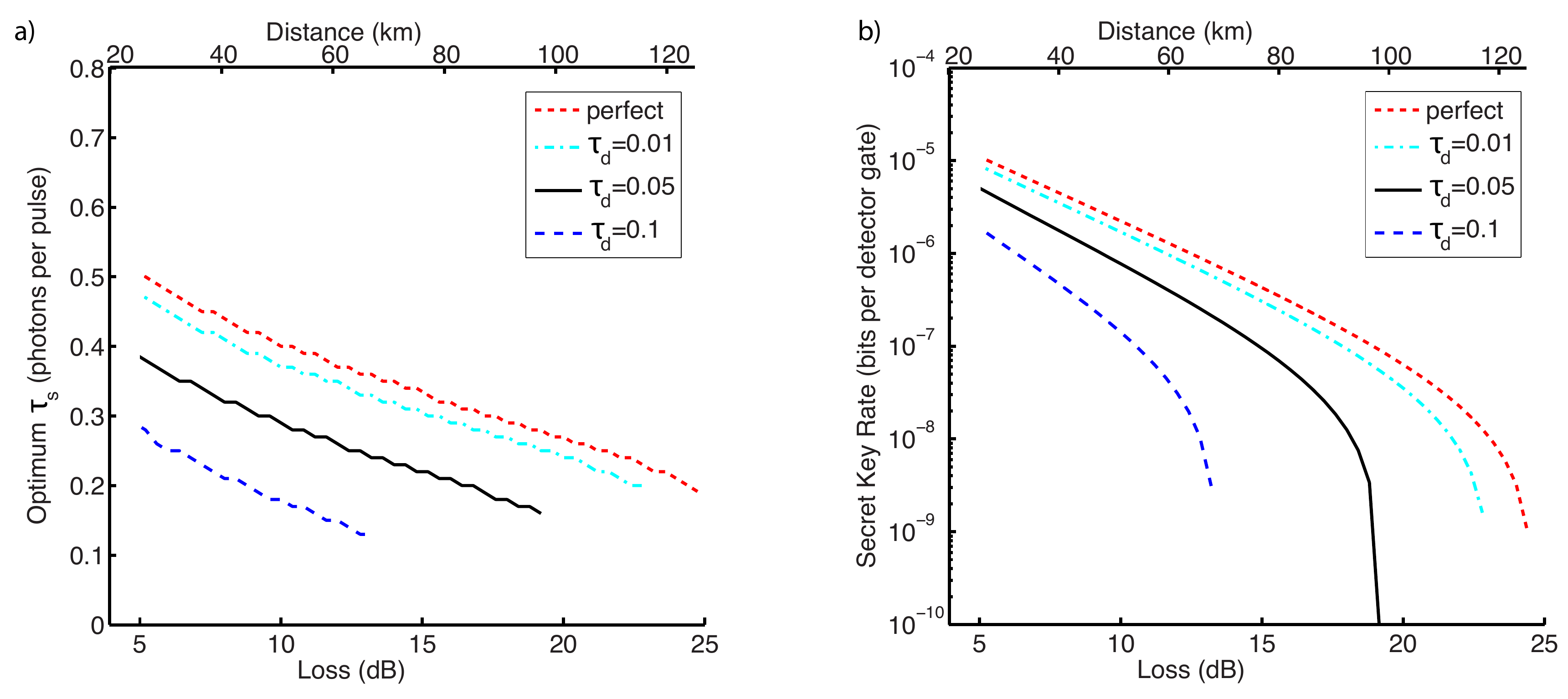}
        \begin{center}
           \caption{\label{fig:optimization} a) Optimum signal state intensity, $\tau_s$, and b) corresponding secret key rate as a function of total loss in dB. The secondary axis shows distances assuming typical loss of 0.2 dB/km in optical fiber without splices. The optimum values for $\mu_s$ for small loss have to be taken with caution as in this regime the model needs to be expanded to higher photon number terms.}
        \end{center}
\end{figure}

In our optimization, we found that, in all cases, $\tau_d = 0.01$ is the optimal decoy intensity.  We attribute this to the fact that $\tau_d$ has a large impact on the tightness of the upper bound on $e_{11}^x$ in Eq.~(\ref{eqn:e11_bound}) (this is due to the fact that all errors in the cases in which both parties sent at least one photon, which increases with $\tau_d$, are attributed to the case in which both parties sent exactly one photon). 
Figure~\ref{fig:optimization} shows, as a function of total loss (or distance), the optimum values of the signal state intensity, $\tau_s$, and the corresponding secret key rate, $S$, for decoy intensities of $\tau_d \in [0.01, 0.05, 0.1]$, as well as for a perfect decoy state protocol (i.e. using values of $Q_{11}^z$ and $e_{11}^x$ computed from the model, as detailed in the preceeding section).

\subsection{Rate-limiting components}\label{sec:limits}
\begin{figure}[h!]
\centering \includegraphics[width=13cm]{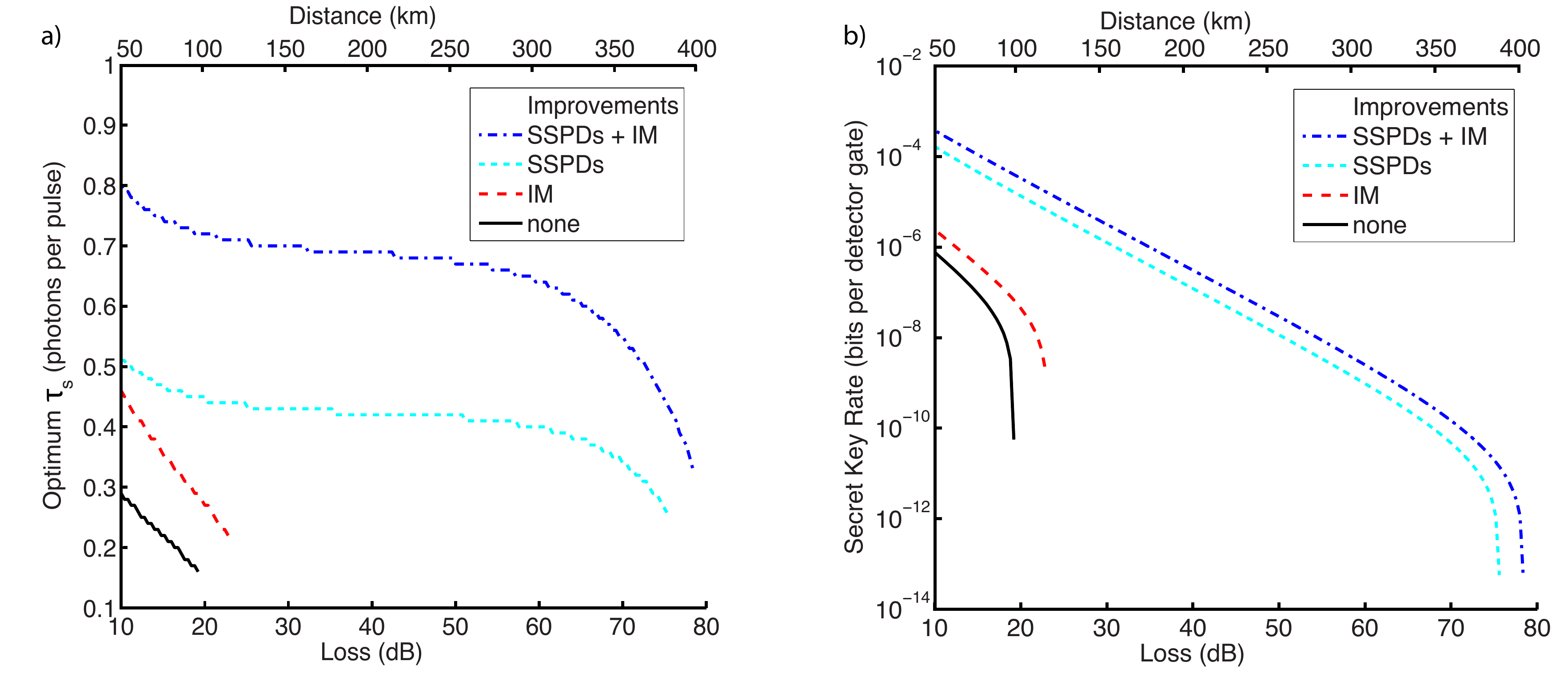}
        \begin{center}
           \caption{\label{fig:ratelimiting} a) Optimum signal state intensity, $\tau_s$, and b) corresponding secret key rate as a function of total loss in dB. The secondary axis shows distances assuming typical loss of 0.2 dB/km in optical fiber without splices. The optimum values for $\mu_s$ for small loss, are not shown as the model needs to be expanded to higher photon number terms in this regime.}
        \end{center}
\end{figure}

Finally, we use our model to simulate the performance of the MDI-QKD protocol given improved components.  We consider two straightforward modifications to the system:  replacing the InGaAs single photon detectors (SPDs) with superconducting single photon detectors (SSPDs)~\cite{Marsili2013}, and improving the intensity modulation (IM).  For various combinations of these improvements, the optimized signal intensities and secret key rates  for $\mu_{d}=0.05$ are shown in Figure~\ref{fig:ratelimiting}.  First, using state-of-the-art SSPDs in~\cite{Marsili2013}, the detection efficiency ($\eta$) is improved from 14.5\% to 93\%, and the dark count probability ($P_{d}$) is reduced by nearly two orders of magnitude.  Furthermore, the mechanisms leading to afterpulsing in InGaAs SPDs are not present in SSPDs (that is $P_{a} = 0$).  This improvement results in a drastic increase in the secret key rate and maximum distance as both the probability of projection onto $\ket{\psi^-}$ and the signal-to-noise-ratio are improved significantly.  Second, imperfections in the intensity modulation system used to create pulses in our implementation contribute significantly to the observed error rates, particularly in the z-basis.  Using commercially-available, state-of-the-art intensity modulators \cite{EOSpace} allow suppressing the background light (represented by $b^{x,z}$ in general quantum state given in Eq.~(\ref{eqn:photon_state})) by an additional ~10-20 dB, corresponding to an extinction ration of 40 dB.  Furthermore, we considered improvements to the driving electronics that reduces ringing in our pulse generation by a factor of 5, bringing the values of $m^{x,z}$ in Eq.~(\ref{eqn:photon_state}) closer to the ideal values.  As seen in Figure~\ref{fig:ratelimiting}, this provides a modest improvement to the secret key rate, both when applied to our existing implementation, and when applied in conjunction with the SSPDs.  Note that in the case of improved detectors and intensity modulation system the optimized $\tau_s$ for small loss (under 10 dB) is likely overestimated due to neglected higher-order terms.

\section{Discussion and conclusion}\label{sec:conclusion}
We have developed a widely applicable model for systems implementing the Measurement-Device-Independent QKD protocol. Our model is based on facts about the experimental setup and takes into account carefully characterized experimental imperfections in sources and measurement devices as well as transmission loss. It is evaluated against data taken with a real, time-bin qubit-based QKD system. The excellent agreement between observed values and predicted data confirms the model. In turn, this allows optimizing mean photon numbers for signal and decoy states and finding rate-limiting components for future improvements. We believe that our model, which is straightforward to generalize to other types of qubit encoding, as well as the detailed description of the characterization of experimental imperfections will be useful to improve QKD beyond its current state of the art. 

To finish, let us emphasize that tests of a model that describes the performance of a QKD system in terms of secret key rates has to happen in a setting in which eavesdropping can be excluded (i.e. within a secure lab and using spooled fibre) -- otherwise, the measured data, which depends on the (unknown) type and amount of eavesdropping, may deviate from the predicted performance and no conclusion about the suitability of the model can be drawn. Interestingly, this implies that neither phase randomization, nor random selection of qubit states or intensities of attenuated laser pulses used to encode qubit states is necessary to test a model, as their presence (or absence) does not impact the measured data. However, it is obvious that these modulations are crucial to ensure the security of a key that is distributed through a hostile environment. We note that in this article, all effects of imperfections in the system on the measured quantities are still attributed to an eavesdropper, and accounted for in the calculation of the secret key rate as well in the optimization of system parameters.

\section*{Acknowledgments}
The authors thank E. Saglamyurek, V. Kiselyov and TeraXion for discussions and technical support, the University of Calgary's Infrastructure Services for providing access to the fiber link between the University's main campus and the Foothills campus, SAIT Polytechnic for providing laboratory space, and acknowledge funding by NSERC, QuantumWorks, General Dynamics Canada, iCORE (now part of Alberta Innovates Technology Futures), CFI, AAET and the Killam Trusts.


\begin{thebibliography}{99}
\bibitem{Gisin2002}{N. Gisin, G. Ribordy, W. Tittel and H. Zbinden, ``Quantum cryptography,'' Rev. Mod. Phys., \textbf{74}, 145-195 (2002).}

\bibitem{Scarani2009}{V. Scarani, H. Bechmann-Pasquinucci, N. J. Cerf, M. Du{\v s}ek, N. L\"utkenhaus and M. Peev, ``The security of practical quantum key disitrbution,'' Rev. Mod. Phys., \textbf{81}, 1301-1350 (2009).}

\bibitem{Dixon2010}{A. Dixon, Z. L. Yuan, J. Dynes, A. W. Sharpe and A. Shields, ``Continuous operation of high bit rate quantum key distribution,'' Appl. Phys. Lett., \textbf{96}, 161102 (2010).}

\bibitem{Stucki2009}{D. Stucki, N. Walenta, F. Vannel, R. T. Thew, N. Gisin, H. Zbinden, S. Gray, C. R. Towery and S. Ten, ``High rate, long-distance quantum key distribution over 250 km of ultra low loss fibres,''  New J. Physics, \textbf{11}, 075003 (2009).}

\bibitem{Schmitt2007}{T. Schmitt-Manderbach, H. Weier, M. F\"urst, R. Ursin, F. Tiefenbacher, T. Scheidl, J. Perdigues, Z. Sodnik, C. Kurtsiefer,  J. G. Rarity, A. Zeilinger and H. Weinfurter, ``Experimental demonstration of free-space decoy-state quantum key distribution over 144 km,'' Phys. Rev. Lett., \textbf{98}, 010504 (2007).}

\bibitem{Masanes2011}{L. Masanes, S. Pironio and A. Ac\'in, ``Secure device-independent quantum key distribution with causally independent measurement devices,'' Nat. Comm., \textbf{2}, 238 (2011).}

\bibitem{Lo2012}{H.-K. Lo, M. Curty and B. Qi, ``Measurement-device-independent quantum key distribution,'' Phys. Rev. Lett., \textbf{108}, 130503 (2012).}

\bibitem{Braunstein2012}{S. L. Braunstein and S. Pirandola, ``Side-channel-free quantum key distribution,'' Phys. Rev. Lett., \textbf{108}, 130502 (2012).}

\bibitem{Rubenok2013}{A. Rubenok, J. A. Slater, P. Chan, I. Lucio-Martinez, W. Tittel, ``Real-world two-photon interference and proof-of-principle quantum key distribution immune to detector attacks,'' Phys. Rev. Lett., \textbf{111}, 130501 (2013).}

\bibitem{Liu2012}{Y. Liu, T.-Y. Chen, L.-J. Wang, H. Liang, G.-L. Shentu, J. Wang, K. Cui, H.-L. Yin, N.-L. Liu, L. Li, X. Ma, J. S. Pelc, M. M. Fejer, Q. Zhang, J.-W. Pan, ``Experimental measurement-device-independent quantum key distribution,'' Phys. Rev. Lett., \textbf{111}, 130502 (2013).}

\bibitem{daSilva2012}{T. F. da Silva, D. Vitoreti, G. B. Xavier, G. P. Tempor\~ao and J. P. von der Weid, ``Proof-of-principle demonstration of measurement device independent QKD using polarization qubits,'' Phys. Rev. A, \textbf{88}, 052303 (2013).}

\bibitem{Tang2013}{Z. Tang, Z. Liao, F. Xu, B. Qi, L. Qian, H.-K. Lo, ``Experimental demonstration of polarization encoding measurement-device-independent quantum key distribution,'' arXiv:1306.6134 [quant-ph].}

\bibitem{Brassard2000}{G. Brassard, N. L{\"u}tkenhaus, T. Mor, B. Sanders, ``Limitation on practical quantum cryptography,'' Phys. Rev. Lett., \textbf{85}, 1330 (2000).}

\bibitem{Hwang2003}{W. Hwang, ``Quantum key distribution with high loss: towards global secure communication,'' Phys. Rev. Lett., \textbf{91}, 057901 (2003).}

\bibitem{Lo2005}{H.-K. Lo, X. Ma and K. Chen, ``Decoy state quantum key distribution,'' Phys. Rev. Lett., \textbf{94}, 230504 (2005).}

\bibitem{Wang2005}{X. Wang, ``Beating the photon-number-splitting attack in practical quantum cryptography,'' Phys. Rev. Lett., \textbf{94}, 230503 (2005).}

\bibitem{Gisin2006}{N. Gisin, S. Fasel, B. Kraus, H. Zbinden and G. Ribordy, ``Trojan-horse attacks on quantum-key-distribution systems,'' Phys. Rev. A, \textbf{73}, 022320 (2006).}

\bibitem{Fung2007}{C.-H. F. Fung, B. Qi, K. Tamaki and H.-K. Lo, ``Phase-remapping attack in practical quantum key distribution systems,'' Phys. Rev. A, \textbf{75}, 032314 (2007).}

\bibitem{Lamas2007}{A. Lamas-Linares and C. Kurtsiefer, ``Breaking a quantum key distribution system through a timing side channel,'' Opt. Express, \textbf{15},  9388-9393 (2007).}

\bibitem{Zhao2008}{Y. Zhao, C.-H. F. Fung, B. Qi, C. Chen and H.-K. Lo, ``Quantum hacking: experimental demonstration of time-shift attack against practical quantum key distribution systems,'' Phys. Rev. A, \textbf{78}, 042333 (2008).}

\bibitem{Wiechers2010}{L. Lydersen, C. Wiechers, C. Wittmann, D. Elser, J. Skaar and V. Makarov, ``Thermal blinding of gated detectors in quantum cryptography,'' Opt. Express, \textbf{18}, 27938-27954 (2010).}

\bibitem{Lydersen2010}{L. Lydersen, C. Wiechers, C. Wittmann, D. Elser, J. Skaar and V. Makarov, ``Hacking commercial quantum cryptography systems by tailored bright illumination,'' Nat. Photonics, \textbf{4}, 686-689 (2010).}

\bibitem{Avoiding1}{Z. L. Yuan, J. F.	Dynes and A. J. Shields, ``Avoiding the blinding attack in QKD,'' Nat. Phot., \textbf{4}, 800-801 (2010).}

\bibitem{Avoiding2}{L. Lydersen, C. Wiechers, C. Wittmann, D. Elser, J. Skaar and V. Makarov, ``Avoiding the blinding attack in QKD,'' Nat. Phot., \textbf{4}, 801 (2010).}

\bibitem{BB84}{C. Bennett and G. Brassard, ``Quantum cryptography: public key distribution and coin tossing,'' Proceedings of IEEE International Conference on Computers Systems and Signal Processing, 175 (1984).}

\bibitem{Wang2013}{X.-B. Wang, ``Three-intensity decoy-state method for device-independent quantum key distribution with basis-dependent errors,'' Phys. Rev. A, \textbf{87}, 012320 (2013). Note that we have corrected a mistake present in Eq.~(17)}.

\bibitem{note}{Note that a pulse does not necessarily contain one single photon. In particular, when considering attenuated light pulses, the number of photons in a pulse will, for example, follow the Poissonian distribution.}

\bibitem{Xu2013}{F. Xu, M. Curty, B. Qi and H.-K. Lo, ``Practical aspects of measurement-device-independent quantum key distribution," New J. Physics, \textbf{15}, 113007 (2013).}

\bibitem{note2}{To the best of our knowledge, this assumption correctly describes all existing experimental implementations. See section \ref{sec:characterization} for more information.}  

\bibitem{note3}{Note that this approximation is, in general, not correct. However, in order to obtain the best performance from a QKD implementation, the noise level should be as low as possible, i.e. $P_n \sim 0$.}

\bibitem{note4}{The separation of photons into ÒgenuineÓ qubit photons and ÒbackgroundÓ photons is somewhat artificial -- as a matter of fact, there is no way to distinguish background photons from real photons. As already stated in section \ref{sec:state}, the distinction is motivated by the need to write down a general expression for all emitted single-photon qubit states using parameters that can be characterized directly through experiments (these measurements are further described below).}

\bibitem{Stucki2001}{D. Stucki, G. Ribordy, A.  Stefanov, H. Zbinden, J. Rarity and T. Wall, ``Photon counting for quantum key distribution with Peltier cooled InGaAs/InP APDs,'' J. of Mod. Opt., \textbf{48}, 1967-1981 (2001).}

\bibitem{Hong1987}{C. K. Hong, Z. Y. Ou and L. Mandel, ``Measurement of subpicosecond time intervals between two photons by interference,'' Phys. Rev. Lett., \textbf{59}, 2044 (1987).}

\bibitem{Mandel83}{L. Mandel, ``Photon interference and correlation effects produced by independent	quantum sources,'' Phys. Rev. A, \textbf{28}, 929 (1983).}

\bibitem{Tamaki11}{K. Tamaki, H.-K. Lo, C.-H. F. Fung and B. Qi, ``Phase encoding schemes for measurement device independent quantum key distribution and basis-dependent flaw," arxiv:1111.3413v4 (2013).}

\bibitem{Sasaki2011}{M. Sasaki \textit{et. al.}, ``Field test of quantum key distribution in the Tokyo QKD network,'' Opt. Express, \textbf{19}, 10387-10409 (2011).}

\bibitem{Marsili2013}{F. Marsili, V. B. Verma, J. A. Stern, S. Harrington, A. E. Lita, T. Gerrits, I. Vayshenker, B. Baek, M. D. Shaw, R. P. Mirin and S. W. Nam, ``Detecting single infrared photons with 93\% system efficiency,'' Nat. Phot., \textbf{7}, 210-214 (2013).}

\bibitem{EOSpace}{For instance, EOSpace sells intensity modulators with 50 dB extinction ratio.}
\end{thebibliography}
\end{document}